\documentclass[conference]{IEEEtran}

\usepackage{graphicx}
\usepackage[cmex10]{amsmath}
\usepackage{amsfonts}
\usepackage{amssymb}
\usepackage{mathrsfs}
\usepackage{bm}
\usepackage{array}
\usepackage{tabularx}
\usepackage{multirow}
\usepackage{color}
\usepackage{textcomp}
\usepackage{tikz}
\usetikzlibrary{mindmap,shadows,arrows,decorations.pathmorphing,backgrounds,positioning,fit,petri}

\usepackage{cite}
\newtheorem{definition}{Definition}

\newtheorem{theorem}{Theorem}
\newtheorem{lemma}{Lemma}

\newtheorem{assumption}{Assumption}

\begin{document}
%
\title{A Dynamic Jamming Game for \\Real-Time Status Updates}

\IEEEoverridecommandlockouts

\author{\IEEEauthorblockN{Yuanzhang Xiao}
\IEEEauthorblockA{Hawaii Center for Advanced Communications\\
University of Hawaii at Manoa\\
Honolulu, HI 96822\\
Email: xyz.xiao@gmail.com} \and \IEEEauthorblockN{Yin Sun}
\IEEEauthorblockA{Department of Electrical and Computer Engineering\\
Auburn University\\
Auburn, AL 36849\\
Email: sunyin02@gmail.com}
\thanks{Yin Sun was supported in part by ONR grant N00014-17-1-2417.}
}


\maketitle

\begin{abstract}
We study timely status updates of a real-time system in an adversarial setting. The system samples a physical process, and sends the samples from the source (e.g., a sensor) to the destination (e.g, a control center) through a channel. For real-time monitoring/control tasks, it is crucial for the system to update the status of the physical process ``timely''. We measure the timeliness of status updates by the time elapsed since the latest update at the destination was generated at the source, and define the time elapsed as \emph{age of information}, or \emph{age} in short. To sabotage the system, an attacker aims to maximize the age by jamming the channel and hence causing delay in status updates. The system aims to minimize the age by judiciously choosing when to sample and send the updates. We model the ongoing repeated interaction between the attacker and the system as a dynamic game. In each stage game, the attacker chooses the jamming time according to the jamming time distribution, and the system responds by delaying the sampling according to the sampling policy. We prove that there exists a unique stationary equilibrium in the game, and provide a complete analytical characterization of the equilibrium. Our results shed lights on how the attacker sabotages the system and how the system should defend against the attacker. 
\end{abstract}

\section{Introduction}

The success of many existing and emerging engineering systems hinges upon the timely status updates of the underlying physical processes. Such systems range from real-time monitoring systems, such as sensor networks monitoring air pollution \cite{air-pollution-2010} and Internet-of-Things (IoT) \cite{IoT-2011}, to networked control systems, such as autonomous vehicles \cite{transportation-2007} and power grids \cite{power-grid-2011}. Timely status updates ensure that the system always has fresh information about the physical process, and that it can monitor and control the process effectively.

The \emph{age of information}, or \emph{age} in short, is becoming the prevailing metric to measure the timeliness of status updates \cite{Liu-1990, Yates-Infocom2012, Elif-ITA2015, Yates-ISIT2015, Ephremides-TIT2016a, Ephremides-TIT2016b, Bedewy2016, Bedewy2017, Yang-ICC2017, Sun2017ISIT, Elif-ISIT2017, Sun-TIT2017, Ephremides-WiOpt2017, SunAoIWorkshop2018, Baknina18, SunAgeIT18}. At any time instant, the age of information is defined as the amount of time elapsed since the most recently received update was generated. A large age indicates that the information about the physical process is outdated, since the most recently received update was generated long time ago. Therefore, the goal is to minimize the age of information. However, minimizing age of information requires careful decisions on when to sample the process and send the samples. Suppose that we have just updated the status. On the one hand, we do not want to update the status immediately, because the new update would be almost the same as the old update, and hence carry little fresh information. On the other hand, we do not want to wait for a long time before updating the status, because the previous update would become outdated before the next update. In summary, we need to make sure that both the information sent and that received are fresh, in order to minimize the age of information \cite{Yates-ISIT2015, Sun-TIT2017}.

While there have been a growing body of works that address the problem of minimizing the age of information in various scenarios \cite{Liu-1990, Yates-Infocom2012, Elif-ITA2015, Yates-ISIT2015, Ephremides-TIT2016a, Ephremides-TIT2016b, Yang-ICC2017, Elif-ISIT2017, Sun-TIT2017}, few works have studied this problem in an adversarial setting \cite{Ephremides-WiOpt2017}. Since timely status updates are of particular importance to real-time monitoring and control systems, the attackers of these systems may specifically target their capability of timely updating the status \cite{Saad-IoT2018,Swami-Computer2016}. Therefore, it is crucial to understand how would an attacker sabotage a real-time monitoring/control system, how should the system defend against the attack, and to what extent the attacker can degrade the timeliness of status update. In this paper, we try to answer these questions in a stylized model.

We formulate a \emph{dynamic game} to model the repeated interaction between a real-time monitoring/control system and its attacker. The system takes samples from a physical process to monitor and sends the samples to a destination through a wireless channel. An attacker jams the channel so that the system cannot send new updates in timely fashion. The attacker aims to find a jamming time distribution to maximize the age of information. After the jamming is over, the system chooses when to sample and send the next update. The system aims to find a sampling policy to minimize the age of information. We summarize the main contributions of this paper as follows.
\begin{itemize}

\item We prove that there exists a unique stationary equilibrium in the game, and that the system cannot further reduce the age at the stationary equilibrium by using more complicated equilibrium strategies.
\item We provide a complete analytical characterization of the equilibrium, which sheds lights on how the system should defend against the attacker. Specifically, the equilibrium sampling policy tries to make the time differences between subsequent update deliveries as equal as possible.
\item We also provide insights on how the attacker should abotage the system. Specifically, the attacker will choose a jamming time distribution with as high variance as possible.
\end{itemize}

\section{Related Works}\label{sec:related}
Most works on age of information do not consider an attacker attacking the system \cite{Liu-1990, Yates-Infocom2012, Elif-ITA2015, Yates-ISIT2015, Ephremides-TIT2016a, Ephremides-TIT2016b, Yang-ICC2017, Elif-ISIT2017, Sun-TIT2017}. The most related work is \cite{Ephremides-WiOpt2017}, where the authors study age of information in an adversarial setting. There are two major differences between our work and \cite{Ephremides-WiOpt2017}. First, our work studies the interaction of the user and the attacker at the MAC layer, while their work \cite{Ephremides-WiOpt2017} studies the interaction in the physical layer. This difference results in different models of actions and more importantly, different models of interaction (e.g., different payoff functions). Second, our work models the interaction as a \emph{dynamic} game, while their work \cite{Ephremides-WiOpt2017} uses a static game model. Therefore, we study the ongoing interaction between the user and the attacker, where the user needs to develop a defense \emph{policy} that specifies a sequence of sampling actions over time. In contrast, their work studies the one-shot interaction, where the user needs to determine a scalar action.

\section{System Model and Problem Formulation}\label{sec:model}

\subsection{System Model}
\begin{figure}
  \centering
  \begin{tikzpicture}[
   output/.style={
      rectangle,minimum size=6mm,rounded corners=3mm,
     very thick,draw=black,
     font=\large},
]
   
   \node at (-0.2,0) [text width=1.2cm,align=center] (physical process) {physical process};
   
   \draw [-latex] (physical process.east) -- (1.5,0);
   \filldraw (1.5,0) circle (0.05);
   \draw [-] (1.5,0) -- (2,0.5);
   \filldraw (2,0.5) circle (0.05);
   \filldraw (2.1,0) circle (0.05);
   \draw [-<] (1.55,0.5) .. controls +(right:0.6) and +(down:0.25) .. (2.1,0.15);
   \node at (1.8, 0.8) {sampler};
   \draw [-] (2.1,0) -- (3,0);
   \draw [-] (3,0) -- (3,0.3);
   \draw [-] (3,0.3) -- (2.8,0.5);
   \draw [-] (2.8,0.5) -- (3.2,0.5);
   \draw [-] (3.2,0.5) -- (3,0.3);
   \node at (3, 0.8) {Tx};
   \draw[thick] (1.1,-0.3) rectangle (3.4,1.1);
   \node at (2.3,-0.6) {sensor};
   
   \draw [-] (5.5,0) -- (5.5,0.3);
   \draw [-] (5.5,0.3) -- (5.3,0.5);
   \draw [-] (5.3,0.5) -- (5.7,0.5);
   \draw [-] (5.7,0.5) -- (5.5,0.3);
   \node at (5.5, 0.8) {Rx};
   \draw [-latex] (5.5,0) -- (6.5,0);
   \node at (7,0.1) [text width=1.2cm,align=center] {monitor control};
   \draw[thick] (5.1,-0.3) rectangle (7.7,1.1);
   \node at (6.4,-0.6) {receiver};
   
   \draw [dashed,-latex] (3.25,0.5) -- (5.3,0.5);
   
   \node at (4.2,2.4) (attacker) {attacker};
   \draw [-latex] (attacker.south) -- (4.2,0.6);
   \node at (5,1.6) {jamming};

  \end{tikzpicture}

\caption{Illustration of the system model.}\label{fig:system-model}
\end{figure}
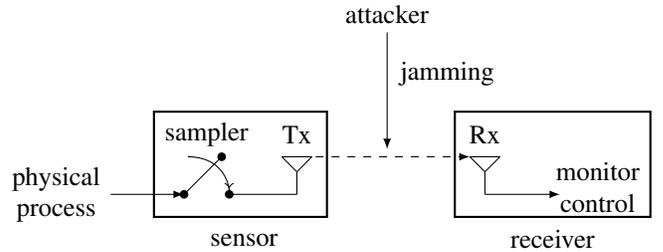

Consider a real-time monitoring system in the presence of an attacker (see Fig.~\ref{fig:system-model} for an illustration). The real-time monitoring system could be a stand-alone system by itself (e.g., sensor networks \cite{air-pollution-2010} and Internet-of-Things \cite{IoT-2011}), or a component of a real-time control system (e.g., autonomous vehicles \cite{transportation-2007} and power grids \cite{power-grid-2011}). The monitoring system consists of a sensor that collects information about some physical process, and a transmitter that sends information to a receiver. The (monitoring) system aims to timely update the information collected, which is crucial for any real-time monitoring or control system. The system achieves timely update of information by judiciously choosing when to sample the physical process and when to transmit the samples. The attacker aims to sabotage the timely status updates of the system, through jamming the wireless channel between the transmitter and the receiver.

Next, we describe the system model in details.\footnote{There are quite a few different modeling choices, such as whether the system employs carrier sensing, which channel the attacker attacks (data or feedback channel), and so on. In this paper, we focus on a model that makes the most sense to us. We leave the comprehensive study of different modeling choices to future works.} 
Since the system updates a small amount of information (e.g., a sensor reading, a control signal) at each time, the duration of each transmission is negligible and assumed to be zero \cite{Elif-ITA2015,Yang-ICC2017,Elif-ISIT2017}. Therefore, although the attacker can monitor the channel and detect a transmission, it is difficult for the attacker to target the transmission (e.g., through colliding with the transmitted packet). Instead, after detecting the transmission of a sample, the attacker can occupy the channel for some amount of time in order to delay the delivery of the next sample. Specifically, to delay the transmission of the $n$-th sample, the attacker jams the channel for a random duration of $A_{n}$. This formulation includes the case when the attacker jams the channel for a deterministic amount of time (by making the distribution a Dirac delta function). We make the following assumption.\footnote{In general, the attacker could choose arbitrarily complex (e.g., time-varying, non-Markovian) distributions of jamming time. 
We would like to consider more general distributions in future works.}

\begin{assumption}
The attacker's jamming time for each transmission is identically and independently distributed (i.i.d.) with probability density function (pdf) $f_A$.
\end{assumption}

We impose the following constraints on the maximum and average jamming time: For all $n=1,2,\ldots$, it holds almost surely that
\begin{eqnarray}\label{eqn:jamming-time-constraints}
0\leq A_n \leq A^{\text{max}}, ~~ \mathbb{E}_{f_A} \{ A_n \} \leq A^{\text{avg}}.
\end{eqnarray}
The maximum jamming time $A^{\text{max}}$ is set up to reduce the chance that the system can locate the attacker. The attacker would like to limit the jamming time to be no longer than $A^{\text{max}}$ to remain elusive. The constraint on the average jamming time comes from the energy constraint of the attacker. Since the attacker's action is the jamming time distribution $f_A$, we translates the constraints in \eqref{eqn:jamming-time-constraints} into constraints on $f_A$ as follows
\begin{eqnarray}\label{eqn:jamming-time-constraints-distribution}
\int_{0}^{A^{\text{max}}} a f_A(a) \text{d}a \leq A^{\text{avg}}, ~\int_{0}^{A^{\text{max}}} f_A(a) \text{d}a = 1.
\end{eqnarray}

The system employs carrier sensing to monitor the channel. Once the channel is idle (i.e., after the attacker stops occupying the channel), the system decides when to sample the physical process and when to transmit the sample. One may wonder why the system does not sample and transmit immediately after the channel is idle? While the system should transmit the sample immediately after sampling (for timely update of information), it may not be optimal to sample immediately after the channel becomes idle. We refer interested readers to \cite[Sec.~I]{Sun-TIT2017} for an example illustrating why the system may want to delay the sampling. We write $D_n$ as the delay of the $n$-th sample after the channel is idle. Denoting the sampling time of the $n$-th sample as $S_n$, we have
\begin{eqnarray}
S_n = S_{n-1} + A_n + D_n, ~\text{for all}~n=1,2,\ldots,
\end{eqnarray}
where we define $S_0 = 0$.

\begin{figure}
  \centering
  \begin{tikzpicture}[
   output/.style={
      rectangle,minimum size=6mm,rounded corners=3mm,
     very thick,draw=black,
     font=\large},
]
   
   \draw [->] (-0.2,0) -- (7,0);
   \node at (7,-0.22) {$t$};
   \draw [->] (0,0) -- (0,3.5);
   \node at (-0.5, 3.5) {$\Delta(t)$};
   
   \draw [thick] (0,0) -- (3,3);
   \draw [thick] (3,3) -- (3,0);
   \draw [|latex - latex] (0,-0.5) -- (1.2,-0.5);
   \node[fill=white] at (0.6, -0.5) {$A_1$};
   \draw [|<->|] (1.2,-0.5) -- (3,-0.5);
   \node[fill=white] at (2.1, -0.5) {$D_1$};
   
   \draw [thick] (3,0) -- (5.5,2.5);
   \draw [thick] (5.5,2.5) -- (5.5,0);
   \draw [|latex-latex] (3,-0.5) -- (4.5,-0.5);
   \node[fill=white] at (3.75, -0.5) {$A_2$};
   \draw [|<->|] (4.5,-0.5) -- (5.5,-0.5);
   \node[fill=white] at (5, -0.5) {$D_2$};
   
   \draw [thick] (5.5,0) -- (6.5, 1);
   
   \filldraw (0,0) circle (0.04);
   \node at (0,-0.22) {$S_0$};
   \filldraw (3,0) circle (0.04);
   \node at (3,-0.22) {$S_1$};
   \filldraw (5.5,0) circle (0.04);
   \node at (5.5,-0.22) {$S_2$};
   
   \node at (0.9, -1.3) {jamming by attacker:};
   \draw [->] (-0.2,-1.8) -- (7,-1.8);
   \node at (7,-2) {$t$};
   \draw[fill=gray] (0,-1.8) rectangle (1.2,-1.6);
   \draw[fill=gray] (3,-1.8) rectangle (4.5,-1.6);
   
   \node at (0.7, -2.2) {sampling by system:};
   \draw [->] (-0.2,-2.9) -- (7,-2.9);
   \node at (7,-3.1) {$t$};
   \draw [-latex] (0,-2.9) -- (0,-2.4);
   \draw [-latex] (3,-2.9) -- (3,-2.4);
   \draw [-latex] (5.5,-2.9) -- (5.5,-2.4);

  \end{tikzpicture}

\caption{Illustration of the system dynamics and evolution of the age of information. $S_n$ is the sampling time of the $n$-th sample. The transmission of the sample is assumed to be instant. $A_n$ is the attacker's jamming time for the $n$-th sample, and $D_n$ is the system's delay of the $n$-th sample after the channel is idle.}\label{fig:AoI}
\end{figure}
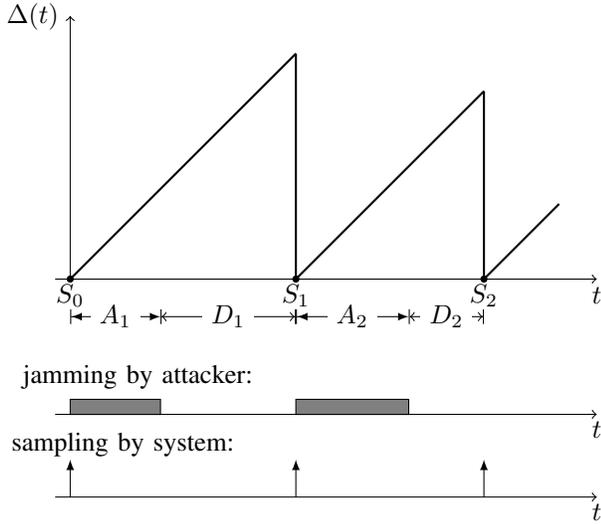

We use the age of information to evaluate the timely update of information \cite{Liu-1990,Yates-Infocom2012}. At any time $t$, the age of information is the time since the last status update of the physical process (i.e., since the last sample). We denote the age of information at time $t$ by $\Delta(t)$, which can be calculated as
\begin{eqnarray}
\Delta(t) = t - S_n, ~\text{for}~t \in [S_n, S_{n+1}).
\end{eqnarray}
The evolution of the age is shown in Fig.~\ref{fig:AoI}.

\subsection{Game Formulation}
Now we formulate the dynamic game between the system and the attacker. The dynamic game consists of an infinite number of stage games. The $n$-th stage game takes place between the transmission of the $(n-1)-$th and the $n$-th samples (i.e., between $S_{n-1}$ and $S_n$). In each stage game $n$, the attacker takes action by choosing the jamming time $A_n$, and the system reacts by choosing delay $D_n$ for the $n$-th sample. In other words, each stage game is a Stackelberg game with the attacker taking action first. Since we have restricted to i.i.d. jamming time distribution, the attacker's action is simply choosing the jamming time distribution $f_A$. We allow the attacker to predict how the system reacts to its action, based on which the attacker will seek its optimal jamming time distribution. 

The system's sampling policy dictates how to choose the delay $D_n$ for each sample. We focuses on causal policies that determine each delay $D_n$ based on information available, which is the historical and current jamming time $A_1,\ldots, A_n$ and the historical delay $D_1,\ldots,D_{n-1}$. We write the set of histories at the beginning of the $n$-th stage game as $\mathcal{H}^n \triangleq [0, A^{\text{max}}]^n \times [0,\infty)^{n-1}$, and the set of all histories as $\mathcal{H} = \cup_{n=1}^\infty \mathcal{H}^n$. Then a causal sampling policy is defined as 
\begin{eqnarray}
& \pi: & \mathcal{H} \rightarrow \Theta\left( [0,\infty) \right),
\end{eqnarray}
where $\Theta\left( [0,\infty) \right)$ is the set of all possible probability distributions with support within $[0,\infty)$. In other words, $\pi(A_1, \ldots, A_n, D_1, \ldots, D_{n-1})$ is the probability distribution of the delay $D_n$.
We write the set of all causal policies as $\Pi$. 

The payoff of interest to both the system and the attacker is the average age of information, defined as
\begin{eqnarray}\label{eqn:average-age}
\overline{\Delta}(f_A, \pi) = \limsup_{T \rightarrow \infty} \frac{1}{T} \mathbb{E}_{f_A}\left[ \int_{0}^{T} \Delta_\pi(t) \text{d}t \right],
\end{eqnarray}
which depends on the attacker's jamming time distribution $f_A$ and the system's sampling policy $\pi$. 

The system aims to minimize the average age $\overline{\Delta}(f_A, \pi)$. Its optimization problem to solve is
\begin{eqnarray}\label{eqn:user-AoI-minimization}
&\displaystyle\min_{\pi \in \Pi} & \overline{\Delta}(f_A, \pi).
\end{eqnarray}

The attacker aims to maximize the average age $\overline{\Delta}(f_A, \pi)$, in anticipation of the system's reaction to its jamming time distribution. Given any jamming time distribution $f_A$, the system reacts by employing a sampling policy that solves \eqref{eqn:user-AoI-minimization}. We write $\mathcal{BR}(f_A)$ as the system's reaction to $f_A$ (i.e., a solution to \eqref{eqn:user-AoI-minimization}). The function $\mathcal{BR}(f_A)$ is also called \emph{best response} in game theoretic term. Note that under the same jamming time distribution $f_A$, the best response $\mathcal{BR}(f_A)$ of the system may not be unique. However, since all best responses result in the same average age, we only need to consider an arbitrary best response.

The attacker's optimization problem can then be written as
\begin{eqnarray}\label{eqn:attacker-AoI-maximization}
&\displaystyle\max_{f_A} & \overline{\Delta}\left[ f_A, \mathcal{BR}(f_A) \right] \\
&s. t.                        & f_A ~\text{satisfies}~ \eqref{eqn:jamming-time-constraints-distribution}. \nonumber
\end{eqnarray}
Note that we use the best response $\mathcal{BR}(f_A)$, instead of an arbitrary policy $\pi$, in the objective function of the attacker. This reflects the fact that the attacker can anticipate the system's best response when taking actions. In other words, each stage game is a Stackelberg game with the attacker being the leader \cite{OsborneRubinstein}. Therefore, the attacker has strategic advantage, because it can predict how the system reacts to its action, and choose its optimal action based on this prediction. By giving the attacker strategic advantage, we can evaluate how well the system can defend in the worst case.

Now we can define the equilibrium of the dynamic game. 
\begin{definition}[Equilibrium] Aa equilibrium is a duple of two functions $\left(f_A^*, \pi^* \right)$ such that
\begin{itemize}
\item the system's sampling policy $\pi^*$ minimizes the average age under the jamming time distribution $f_A^*$, namely $\pi^*$ is a solution to \eqref{eqn:user-AoI-minimization}.
\item the attacker's jamming time distribution $f_A^*$ maximizes the average age in anticipation of the system's best response, namely $f_A^*$ is a solution to \eqref{eqn:attacker-AoI-maximization}.
\end{itemize}
\end{definition}
The definition of equilibrium ensures that neither the system nor the attacker could deviate from the equilibrium and get more favorable outcomes.

\section{Main Result}\label{sec:result}
We give a complete characterization of equilibria in this section. Note that both optimization problems of the system and the attacker in \eqref{eqn:user-AoI-minimization} and \eqref{eqn:attacker-AoI-maximization} are complicated functional optimization problems. We first argue that we can simplify the problem by restricting our attention to stationary deterministic sampling policies. We then define the corresponding stationary equilibrium, prove that the stationary equilibrium is unique, and provide analytical expressions of the equilibrium jamming time distribution and the equilibrium sampling policy.

\subsection{Restriction to Stationary Deterministic Sampling Policies}
We consider \emph{stationary deterministic} sampling policies, a special class of causal policies defined as follows.
\begin{definition}[Stationary Deterministic Policy]
A causal policy is said to be \emph{stationary deterministic}, if there exists a deterministic function $u: [0, A^{\text{max}}] \rightarrow [0, \infty)$, such that
for all $(A_1, \ldots, A_n, D_1, \ldots, D_{n-1}) \in \mathcal{H}^n$, we have
\begin{eqnarray}\label{eqn:definition-stationary-deterministic-policy}
U_n = u(A_n).
\end{eqnarray}
\end{definition}
A stationary deterministic policy chooses the delay $D_n$ as a fixed, deterministic function of the jamming time $A_n$, independent of the historical jamming time $A_1,\ldots,A_{n-1}$ and historical delay $D_1,\ldots,D_{n-1}$.

Although the system could choose any causal policy $\pi \in \Pi$, the following lemma suggests that given any jamming distribution $f_A$, there exists a \emph{unique} stationary deterministic sampling policy that is optimal among all causal policies.
\begin{lemma}\label{lemma:stationary-deterministic-policy-is-optimal}
Given any jamming time distribution $f_A$, there exists a unique stationary deterministic sampling policy that is optimal among all causal policies.
\end{lemma}
\begin{IEEEproof}
For a given jamming distribution $f_A$, the age minimization problem of the system is equivalent to the one in \cite{Sun-TIT2017}. Therefore, we can use \cite[Theorem~4]{Sun-TIT2017}. The results in Lemma~\ref{lemma:stationary-deterministic-policy-is-optimal} directly come from \cite[Theorem~4]{Sun-TIT2017}.
\end{IEEEproof}

Lemma~\ref{lemma:stationary-deterministic-policy-is-optimal} allows us to restrict to stationary deterministic sampling policies. According to \eqref{eqn:definition-stationary-deterministic-policy}, a stationary deterministic policy is completely specified by the deterministic function $u$. Hence, we will refer to a stationary deterministic policy as $u$.

Since we restrict to stationary deterministic sampling policies, we focus on \emph{stationary} equilibrium defined as follows.
\begin{definition}[Stationary Equilibrium] A stationary equilibrium is an equilibrium $\left(f_A^*, \pi^* \right)$ where the sampling policy $\pi^*$ is stationary deterministic.
\end{definition}

We also write a stationary equilibrium as $\left(f_A^*, u^* \right)$.

\subsection{Characterization of Equilibria}

Now we give a complete characterization of stationary equilibrium.
\begin{theorem}\label{theorem:characterization-of-equilibrium}
There is a unique stationary equilibrium, where the attacker's equilibrium jamming time distribution is
\begin{eqnarray}
f_A^*(a) = \left( 1 - \frac{ A^{\text{avg}} } { A^{\text{max}} } \right) \cdot \delta(a) + \frac{ A^{\text{avg}} } { A^{\text{max}} } \cdot \delta(a - A^{\text{max}}),
\end{eqnarray}
where $\delta(\cdot)$ is the Dirac delta function, and the system's equilibrium sampling policy is $D_n = u^*(A_n)$, where the function $u^*$ is given by
\begin{eqnarray}
u^*(a) = ( \beta^* - a )^+, ~\text{for all}~a \in [0, A^{\text{max}}],
\end{eqnarray}
where $(\cdot)^+ \triangleq \max\left\{ \cdot, 0 \right\}$, and $\beta^*$ is a constant given by
\begin{eqnarray}
\beta^* \triangleq \frac{ \sqrt{A^{\text{avg}}} } { \sqrt{A^{\text{max}}} + \sqrt{A^{\text{avg}}} } \cdot A^{\text{max}}.
\end{eqnarray}
\end{theorem}
\begin{IEEEproof}
See Section~\ref{sec:proof}.
\end{IEEEproof}

Our result provides analytical characterization of the equilibrium, and hence shed lights on the equilibrium behavior of the system and the attacker. First, the system's policy can be viewed as a ``water-filling'' policy with $\beta$ as the water level. The purpose of water-filling is to make the intervals between receipts of consecutive samples (i.e., $S_n - S_{n_1}$) as equal as possible. When the attacker's jamming time $A_n=a$ is short, the system chooses a larger delay $D_n = u(a)$, so that the new transmission carries ``fresh'' information. More specifically, if the jamming time is zero, and if the system chooses zero delay, the new transmission contains exactly the same information as the previous transmission, and hence the new transmission is ``wasted''. On the contrary, the system chooses a smaller delay if the attacker's jamming time is long, and does not delay the transmission at all if the jamming time is above the threshold $\beta$. This ensures that the received information is not outdated by a large delay in the transmission. With such a water-filling policy, the system strikes the balance between sending fresh enough information (by not letting the delay too small) and receiving fresh enough information (by not letting the delay too large).

Since the system wishes to keep the time between receipts of consecutive samples equal, the attacker tries to make it difficult for the system to achieve this. The worst jamming distribution for the attacker is a fixed deterministic jamming time, against which the system can easily make sure that the updates are received periodically. Therefore, the best jamming distribution for the attacker should have high variance, where the realized jamming time is either the largest ($A^{\text{max}}$) or zero. This is exactly the equilibrium jamming time distribution specified in our theorem.

\section{Proof of Main Result}\label{sec:proof}

In this section, we prove our main results in Theorem~\ref{theorem:characterization-of-equilibrium}. We first derive the best response sampling policy $\mathcal{BR}(f_A)$, and then solve for the equilibrium jamming time distribution $f_A^*$. Based on these results, we can obtain the equilibrium sampling policy $u^* = \mathcal{BR}(f_A^*)$.

\subsection{Best Response Sampling Policy}
First, we derive the best response sampling policy $\mathcal{BR}(f_A)$ under any jamming time distribution $f_A$.

\begin{lemma}\label{lemma:best-response-sampling}
For any jamming time distribution $f_A$ that satisfies \eqref{eqn:jamming-time-constraints-distribution}, the best response sampling policy is $\mathcal{BR}(f_A) = u$, where
\begin{eqnarray}
u(a) = (\beta - a)^+,
\end{eqnarray}
with $\beta$ satisfying
\begin{eqnarray}
\beta = \frac{ \mathbb{E}_{f_A}\left\{ \left[ A + u(A) \right]^2 \right\} } { 2 \cdot \mathbb{E}_{f_A}\left[ A + u(A) \right] }.
\end{eqnarray}
\end{lemma}
\begin{IEEEproof}
We have proved in Lemma~\ref{lemma:stationary-deterministic-policy-is-optimal} that the best response sampling policy is characterized by \cite[Theorem~4]{Sun-TIT2017}. The results directly follow from \cite[Theorem~4]{Sun-TIT2017}.
\end{IEEEproof}

\subsection{Equilibrium Jamming Time Distribution}
Next, we derive the equilibrium jamming time distribution $f_A^*$. We first simplify the attacker's optimization problem \eqref{eqn:attacker-AoI-maximization} by the following lemma.
\begin{lemma}\label{lemma:attacker-AoI-maximization-simplified}
The attacker's optimization problem \eqref{eqn:attacker-AoI-maximization} is equivalent to
\begin{eqnarray}\label{eqn:attacker-AoI-maximization-simplified}
& \displaystyle\max_{f_A} & \frac{ \int_{0}^{A^{\text{max}}} \left[ \max\{ \beta , a \} \right]^2 f_A(a) \text{d}a }
							          { \int_{0}^{A^{\text{max}}} \max\{ \beta , a \} f_A(a) \text{d}a } \\
& s. t.                                & \beta = \frac{1}{2} \cdot \frac{ \int_{0}^{A^{\text{max}}} \left[ \max\{ \beta , a \} \right]^2 f_A(a) \text{d}a }
							         	              { \int_{0}^{A^{\text{max}}} \max\{ \beta , a \} f_A(a) \text{d}a } \nonumber \\
&                                                        & f_A ~\text{satisfies}~ \eqref{eqn:jamming-time-constraints-distribution}. \nonumber
\end{eqnarray}
\end{lemma}
\begin{IEEEproof}
See Subsection A of Appendix. 
\end{IEEEproof}

We verify that the equilibrium jamming time distribution $f_A^*$ in Theorem~\ref{theorem:characterization-of-equilibrium} is a feasible solution to \eqref{eqn:attacker-AoI-maximization-simplified}.
\begin{lemma}\label{lemma:feasibility}
The equilibrium jamming time distribution $f_A^*$ in Theorem~\ref{theorem:characterization-of-equilibrium} is a feasible solution to \eqref{eqn:attacker-AoI-maximization-simplified}. Moreover, under $f_A^*$, the unique $\beta$ that satisfies the first constraint of \eqref{eqn:attacker-AoI-maximization-simplified} is the $\beta^*$ defined in Theorem~\ref{theorem:characterization-of-equilibrium}.
\end{lemma}
\begin{IEEEproof}
See Subsection B of Appendix. 
\end{IEEEproof}

We need two properties of the objective function of \eqref{eqn:attacker-AoI-maximization-simplified} to prove that the feasible solution $f_A^*$ is indeed the unique optimal solution. We write the objective of \eqref{eqn:attacker-AoI-maximization-simplified} as a function of $\beta$ and $f_A$ as follows:
\begin{eqnarray}
g(\beta, f_A) \triangleq \frac{ \int_{0}^{A^{\text{max}}} \left[ \max\{ \beta , a \} \right]^2 f_A(a) \text{d}a }
					 { \int_{0}^{A^{\text{max}}} \max\{ \beta , a \} f_A(a) \text{d}a }.
\end{eqnarray}
Note that the argument $\beta$ does not need to satisfy the first constraint in \eqref{eqn:attacker-AoI-maximization-simplified}.

The first useful property of the function $g(\beta, f_A)$ is that when the jamming time distribution is equal to the equilibrium distribution (i.e., when $f_A = f_A^*$), the largest $\beta$ that satisfies $\beta = \frac{1}{2} g(\beta, f_A^*)$ is $\beta^*$.

\begin{lemma}\label{lemma:largest-beta}
The largest $\beta$ that satisfies $\beta = \frac{1}{2} g(\beta, f_A^*)$ is $\beta^*$. In addition, we have $\beta > \frac{1}{2} g(\beta, f_A^*)$ for any $\beta > \beta^*$.
\end{lemma}
\begin{IEEEproof}
See Subsection C of Appendix. 
\end{IEEEproof}

The second useful property of the function $g(\beta, f_A)$ is that for any $\beta$, the equilibrium distribution $f_A^*$ maximizes $g(\beta, f_A)$ among all distributions that satisfy \eqref{eqn:jamming-time-constraints-distribution}.
\begin{lemma}\label{lemma:worst-distribution}
Given any $\beta$, the equilibrium distribution $f_A^*$ is a solution to the following problem:
\begin{eqnarray}
& \displaystyle\max_{f_A} & \frac{ \int_{0}^{A^{\text{max}}} \left[ \max\{ \beta , a \} \right]^2 f_A(a) \text{d}a }
							          { \int_{0}^{A^{\text{max}}} \max\{ \beta , a \} f_A(a) \text{d}a } \\
& s. t.                                & f_A ~\text{satisfies}~ \eqref{eqn:jamming-time-constraints-distribution}. \nonumber
\end{eqnarray}
\end{lemma}
\begin{IEEEproof}
See Subsection D of Appendix. 
\end{IEEEproof}

Base on Lemma~\ref{lemma:largest-beta} and Lemma~\ref{lemma:worst-distribution}, we can prove that $f_A^*$ is the unique optimal solution to \eqref{eqn:attacker-AoI-maximization-simplified} by contradiction. Suppose that there is a better solution $\tilde{f}_A$ such that $g(\tilde{\beta}, \tilde{f}_A) > g(\beta^*, f_A^*)$. Since $\tilde{f}_A$ is feasible, we have $\tilde{\beta} = \frac{1}{2} g(\tilde{\beta}, \tilde{f}_A)$, which implies $\tilde{\beta} > \frac{1}{2} g(\beta^*, f_A^*) = \beta$. Base on Lemma~\ref{lemma:worst-distribution}, we have
\begin{eqnarray}
g(\tilde{\beta}, \tilde{f}_A) \leq g(\tilde{\beta}, f_A^*).
\end{eqnarray}
Base on Lemma~\ref{lemma:largest-beta} and the fact that $\tilde{\beta} > \beta$, we have
\begin{eqnarray}
g(\tilde{\beta}, f_A^*) < 2 \tilde{\beta}.
\end{eqnarray}
Therefore, we have $g(\tilde{\beta}, \tilde{f}_A) < 2 \tilde{\beta}$, which violates the first constraint of \eqref{eqn:attacker-AoI-maximization-simplified}. This contradiction implies that 
$f_A^*$ is the unique optimal solution to \eqref{eqn:attacker-AoI-maximization-simplified}.

\subsection{Equilibrium Sampling Policy}
It was proved in Lemma~\ref{lemma:feasibility} that given equilibrium jamming time distribution $f_A^*$, the $\beta^*$ specified in Theorem~\ref{theorem:characterization-of-equilibrium} is the unique $\beta$ in the best response to $f_A^*$. This concludes the proof.

\section{Simulation Results}\label{sec:simulation}

\begin{figure}
\centering
\includegraphics[clip, trim={1.1cm 6.5cm 1.2cm 6.5cm}, width=6cm]{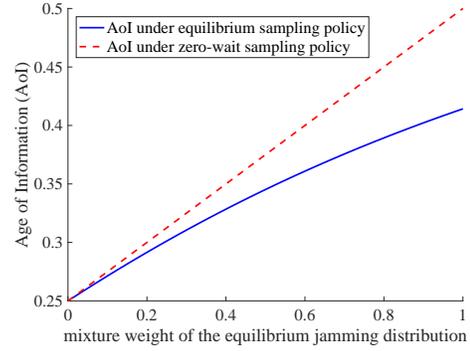}
\caption{Age of information under the equilibrium and zero-wait sampling policies when the mixture jamming distribution changes from the deterministic jamming time to the equilibrium jamming time distribution.}
\label{fig:equilibrium-versus-zero-wait-under-mixture}
\end{figure}

We have argued that the equilibrium jamming time distribution has a large variance and maximizes the age, while a deterministic jamming time is worst in maximizing the age. Now we consider a jamming time distribution that is the mixture of the two distribution:
$$
f_A^\alpha = \alpha \cdot f_A^* + (1-\alpha) \cdot f_A^D,
$$
where $\alpha \in [0,1]$ is the weight of the equilibrium jamming time distribution and $f_A^D$ is a deterministic jamming time with
$$
f_A^D(a) = \delta\left( a - A^{\text{avg}} \right).
$$
We choose the deterministic jamming time to be $A^{\text{avg}}$ so that the mixture distribution has a constant mean value.

Under the mixture distribution, we compute the age under the equilibrium sampling policy defined in \eqref{lemma:best-response-sampling}, and that under the zero-wait sampling policy (i.e., samples immediately after the jamming). Fig.~\ref{fig:equilibrium-versus-zero-wait-under-mixture} shows the ages versus the weight $\alpha$. 

First, we can see that the equilibrium jamming distribution $f_A^*$ is indeed better for the attacker, since the ages in both cases increase as the mixture distribution becomes closer to the equilibrium jamming distribution. Second, we can see that the equilibrium sampling policy achieves smaller ages than the zero-wait policy.

\section{Conclusion}\label{sec:conclusion}
We studied a dynamic game between a real-time system and an attacker. The system samples a physical process and sends the samples through a wireless channel, in order to perform some monitoring and control tasks. The attacker jams the channel to delay the status updates. We used the age of information as the performance metric to measure the timeliness of the status updates. We proved that there exists a unique stationary equilibrium in this game, and characterized the equilibrium analytically. Our results indicate that the attacker chooses a jamming time distribution with high variance, while the system chooses a sampling policy that results in low variance in the time between the receipts of two consecutive updates.

\section*{Appendix}

\subsection{Proof of Lemma~\ref{lemma:attacker-AoI-maximization-simplified}}\label{proof:attacker-AoI-maximization-simplified}
We first calculate the average age $\overline{\Delta}(f_A, u)$ defined in \eqref{eqn:average-age}. Since the jamming time distribution is i.i.d., and since the sampling policy is stationary deterministic, the average age $\overline{\Delta}(f_A, u)$ over the infinite time horizon is equal to the average age of each stage game. Therefore, we have
\begin{eqnarray}\label{eqn:average-age-simplified}
\overline{\Delta}(f_A, u) = \frac{ \mathbb{E}_{f_A}\left[ \left( a + u(a) \right)^2 \right] }
					      { \mathbb{E}_{f_A}\left[ a + u(a) \right] }.
\end{eqnarray}
We can further simplify \eqref{eqn:average-age-simplified} when $u = \mathcal{BR}(f_A)$ (i.e., when $u$ is best response), by observing that
\begin{eqnarray}
a + (\beta - a)^+ = \max\{ \beta, a \}.
\end{eqnarray}
Hence, we have
\begin{eqnarray}\label{eqn:average-age-simplified-2}
\overline{\Delta}\left[ f_A, \mathcal{BR}(f_A) \right] = \frac{ \int_{0}^{A^{\text{max}}} \left[ \max\{ \beta , a \} \right]^2 f_A(a) \text{d}a }
					                                   { \int_{0}^{A^{\text{max}}} \max\{ \beta , a \} f_A(a) \text{d}a }.
\end{eqnarray}
At this point, the attacker's optimization problem \eqref{eqn:attacker-AoI-maximization} can be rewritten as
\begin{eqnarray}
& \displaystyle\max_{f_A} & \frac{ \int_{0}^{A^{\text{max}}} \left[ \max\{ \beta , a \} \right]^2 f_A(a) \text{d}a }
							          { \int_{0}^{A^{\text{max}}} \max\{ \beta , a \} f_A(a) \text{d}a } \\
& s. t.                                & \beta = \frac{1}{2} \cdot \frac{ \int_{0}^{A^{\text{max}}} \left[ \max\{ \beta , a \} \right]^2 f_A(a) \text{d}a }
							         	              { \int_{0}^{A^{\text{max}}} \max\{ \beta , a \} f_A(a) \text{d}a } \nonumber \\
&                                                        & f_A ~\text{satisfies}~ \eqref{eqn:jamming-time-constraints-distribution}. \nonumber
\end{eqnarray}
Note that we need the first constraint to ensure that the system is choosing the best response sampling policy.

\subsection{Proof of Lemnma~\ref{lemma:feasibility}}
Since we have 
\begin{eqnarray}
\text{supp}(f_A^*) = \{ 0, A^{\text{max}} \} \subset [ 0, A^{\text{max}} ],
\end{eqnarray}
and
\begin{eqnarray*}
\int_{0}^{A^{\text{max}}} a f_A^*(a) \text{d}a = 
\left( 1 - \frac{ A^{\text{avg}} } { A^{\text{max}} } \right) \cdot 0 + \frac{ A^{\text{avg}} } { A^{\text{max}} } \cdot A^{\text{max}} 
= A^{\text{avg}},
\end{eqnarray*}
we know that $f_A^*$ satisfies the second constraint of \eqref{eqn:attacker-AoI-maximization-simplified}.

Now it remains to show that there exists a unique $\beta = \beta^*$ that satisfies
\begin{eqnarray}\label{eqn:first-constraint}
\beta = \frac{1}{2} \cdot \frac{ \int_{0}^{A^{\text{max}}} \left[ \max\{ \beta , a \} \right]^2 f_A^*(a) \text{d}a }
					   { \int_{0}^{A^{\text{max}}} \max\{ \beta , a \} f_A^*(a) \text{d}a }.
\end{eqnarray}
Suppose that $\beta = 0$. Then \eqref{eqn:first-constraint} reduces to
\begin{eqnarray}
0 = \frac{1}{2} \frac{ (A^{\text{max}})^2 \cdot ( A^{\text{avg}} / A^{\text{max}} ) } 
			     { (A^{\text{max}})    \cdot ( A^{\text{avg}} / A^{\text{max}} ) }
   = \frac{ A^{\text{max}} } { 2 },
\end{eqnarray}
which is impossible. Suppose that $\beta \geq A^{\text{max}}$. Then \eqref{eqn:first-constraint} reduces to
\begin{eqnarray}
A^{\text{max}} &=& \frac{1}{2} \frac{ (A^{\text{max}})^2 ( 1 - A^{\text{avg}} / A^{\text{max}} ) + (A^{\text{max}})^2 ( A^{\text{avg}} / A^{\text{max}} ) } 
			     { (A^{\text{max}}) ( 1 - A^{\text{avg}} / A^{\text{max}} ) + (A^{\text{max}}) ( A^{\text{avg}} / A^{\text{max}} ) } \nonumber \\
                        &=& \frac{ A^{\text{max}} } { 2 },
\end{eqnarray}
which is impossible. Therefore, we must have $\beta \in ( 0 , A^{\text{max}} )$. Then \eqref{eqn:first-constraint} reduces to
\begin{eqnarray}
&                       & \beta = \frac{1}{2} \frac{ \beta^2 ( 1 - A^{\text{avg}} / A^{\text{max}} ) + (A^{\text{max}})^2 ( A^{\text{avg}} / A^{\text{max}} ) } 
			               { \beta ( 1 - A^{\text{avg}} / A^{\text{max}} ) + (A^{\text{max}}) ( A^{\text{avg}} / A^{\text{max}} ) } \nonumber \\
&\Leftrightarrow& \left(1 - \frac{A^{\text{avg}}}{A^{\text{max}}} \right) \beta^2 + 2 A^{\text{avg}} \beta - A^{\text{avg}} A^{\text{max}} = 0.
\end{eqnarray}
It is not difficult to verify that the only positive solution to the above equation is
\begin{eqnarray}
\beta = \frac{ \sqrt{A^{\text{avg}}} } { \sqrt{A^{\text{max}}} + \sqrt{A^{\text{avg}}} } \cdot A^{\text{max}}.
\end{eqnarray}
Hence, under $f_A^*$, the unique $\beta$ that satisfies the first constraint of \eqref{eqn:attacker-AoI-maximization-simplified} is the $\beta^*$ defined in Theorem~\ref{theorem:characterization-of-equilibrium}.

\subsection{Proof of Lemma~\ref{lemma:largest-beta}}
In Lemma~\ref{lemma:feasibility}, we have shown that under the equilibrium distribution $f_A^*$, $\beta^*$ satisfies the first constraint of \eqref{eqn:attacker-AoI-maximization-simplified}, namely $\beta^* = \frac{1}{2} g(\beta^*, f_A^*)$. Now we show that $\beta > \frac{1}{2} g(\beta, f_A^*)$ for any $\beta > \beta^*$.

The derivative of $\beta - \frac{1}{2} g(\beta, f_A^*)$ with respect to $\beta$ is
\begin{eqnarray}
&  & \frac{ \partial \left[ \beta - \frac{1}{2} g(\beta, f_A^*) \right] } { \partial \beta } \\
&=& 1 - \frac{1}{2} \cdot 
             \frac{ \left( 1 - \frac{A^{\text{avg}}}{A^{\text{max}}} \right) \beta^2 + 2 A^{\text{avg}} \beta - A^{\text{avg}} A^{\text{max}} }
                    { \left( 1 - \frac{A^{\text{avg}}}{A^{\text{max}}} \right) \beta^2 + 2 A^{\text{avg}} \beta + (A^{\text{avg}})^2 / \left( 1 - \frac{A^{\text{avg}}}{A^{\text{max}}} \right) } \nonumber \\
&>&  1 - \frac{1}{2} = \frac{1}{2} > 0. \nonumber
\end{eqnarray}

Since $\beta^* - \frac{1}{2} g(\beta^*, f_A^*) = 0$, and since $\frac{ \partial \left[ \beta - \frac{1}{2} g(\beta, f_A^*) \right] } { \partial \beta } > 0$, we have $\beta > \frac{1}{2} g(\beta, f_A^*)$ for any $\beta > \beta^*$.

\subsection{Proof of Lemma~\ref{lemma:worst-distribution}}
The objective function is a ratio of the weighted sum of $[\max\{\beta,a\}]^2$ to the weighted sum of $\max\{\beta,a\}$, where the weight is determined by $f_A$. Since $\frac{ [\max\{\beta,a\}]^2 } { \max\{\beta,a\} }$ is non-decreasing in $a$, we should assign the largest possible weight when $a = A^{\text{max}}$. However, the distribution needs to satisfy the average jamming time condition. The equilibrium distribution $f_A^*$ has the largest $f_A^*( A^{\text{max}} )$ among all the distributions that satisfy \eqref{eqn:jamming-time-constraints-distribution}.

\bibliographystyle{IEEEtran}
\bibliography{IEEEabrv,xyz_AoI}
%
%
%
%

\end{document}